\documentclass[journal, onecolumn]{IEEEtran}
\usepackage{amsmath,amssymb,amsthm,amstext}
\usepackage{t1enc}
\usepackage[utf8]{inputenc}
\usepackage{graphicx,subcaption}
\usepackage{color}
\usepackage{latexsym}
\usepackage{multicol,multirow}
\usepackage{verbatim}
\usepackage{hyperref}

\DeclareMathOperator{\Esp}{\mathrm I\!\mathrm E}

\DeclareMathOperator{\Prob}{\mathrm I\!\mathrm P}

\newtheorem{theorem}[]{Theorem}
\newtheorem{corollary}[]{Corollary}
\newtheorem{lemma}[]{Lemma}

\newenvironment{remark}[1][remark]{\begin{trivlist}
  \item[\hskip \labelsep {\textsc{#1}}]}{\end{trivlist}}

\begin{document}
\title{Irregular Repetition Slotted Aloha with Multiuser Detection: A
  Density Evolution Analysis}
\author{Manuel Fernández-Veiga, M.E. Sousa-Vieira, Ana Fernández-Vilas,  Rebeca P. Díaz-Redondo
\thanks{mveiga@det.uvigo.es, estela@det.uvigo.es, avilas@det.uvigo.es, rebeca@det.uvigo.es; atlanTTic, Universidade de Vigo; Vigo, 36310, Spain.}}


\maketitle

\begin{abstract}
  Irregular repetition slotted Aloha (IRSA) has shown significant
  advantages as a modern technique for uncoordinated random access
  with massive number of users due to its capability of achieving
  theoretically a throughput of $1$ packet per slot. When the receiver
  has also the multi-packet reception of multi-user (MUD) detection
  property, by applying successive interference cancellation, IRSA also
  obtains very low packet loss probabilities at low traffic loads, but
  is unable in general to achieve a normalized throughput close to the
  $1$. In this paper, we reconsider the case of IRSA with $k$-MUD
  receivers and derive the general density evolution equations for the
  non-asymptotic analysis of the packet loss rate, for arbitrary frame
  lengths and two variants of the first slot used for
  transmission. Next, using the potential function, we give new
  capacity bounds on the capacity of the system, showing the threshold
  arrival rate for zero decoding error probability. Our numerical
  results illustrate performance in terms of throughput and average
  delay for $k$-MUD IRSA with finite memory at the receiver, and also
  with bounded maximum delay.
\end{abstract}


\section{Introduction}
\label{sec:Introduction}

Simple and scalable protocols for massive random access are essential
for supporting large deployments of the Internet of Things (IoT) and
for large-scale machine-type communications (mMTC), as in Industry 4.0
and other specialised 5G/6G
verticals~\cite{Aceto2019,LeyvaMayorga2019}, which are starting to
show an explosive commercial growth. In this context, in order to
satisfy the stringent key performance targets of this type of systems
in which an unknown fraction of devices out of a large collection are
actively transmitting packets, the design of multiple access
techniques for the next generation wireless networks has become a key
enabler technology~\cite{Liu2022}. Uncoordinated transmissions on a
multiple access channel (MAC) with random access and retransmissions,
like in classical ALOHA~\cite{Abramson1977} and variants, have very
poor performance in throughput, spectral (SE) and energy efficiency
(EE) when they are used in massive access conditions, since collisions
waste a substantial amount of physical resources (resource blocks and
energy) and may introduce large delays. This has motivated the
investigation of \emph{efficient} variants of ALOHA which keep its
simplicity of operations yet can sustain high throughput and EE for
IoT and mMTC applications, i.e., for low-power, low-complexity
devices. Some of these grant-free novel MAC protocols have become part
of standards such as NB-IoT or ETSI DVB-RCS2, and are receiving much
attention as solution for next-generation MAC techniques in 5G and 6G
networks~\cite{Liu2022}.

A modern approach toward an efficient use of channel resources in
massive MAC is to use non-orthogonal encoded transmission of packets
combined with some form of successive interference cancellation (SIC)
at the receiver~\cite{Casini2007,Liva2011,RioHerrero2014}. In
irregular repetition slotted Aloha (IRSA), the user devices transmit
several replicas of each packet (an elementary form of coding by
replication) in randomly selected slots. The receiver then separates
the simultaneous transmissions in a given slot applying SIC: starting
with some slot in which a single transmission has occurred, the
receiver iteratively looks back in the past slots and attempts to
cancel out the interference in those slots with collisions where a
replica of the decoded packet was also transmitted. SIC removes the
last decoded packet from the superposition of signals received at
another slot, and this eventually enables the effective decoding of
any of the replicas of other packets. The procedure can continue in
the same way until no further opportunity for discovery and SIC-based
decoding is detected or until all packets have been decoded. IRSA can
drastically improve performance of classical
ALOHA~\cite{Meloni2012,Sandgren2017}, achieving throughput close to
$1$ and high EE for up to hundreds of active devices
(see~\cite{Polyanskiy2017} for a modern information-theoretic
formulation of massive random access). The time diversity of the
packet replicas, and the system performance, is increased if each
device follows a variable number of repetitions for its packets drawn
from a probability distribution, which can be optimized.

Coded slotted ALOHA has been extensively studied in the literature in
the asymptotic regime~\cite{RioHerrero2014,Sandgren2017} (where the
packet replicas are spread over a large interval) and in the
finite-length regime~\cite{Amat2018,Ivanov2015,Ivanov2017}. Extensions
of coded ALOHA, and in particular of irregular repetition coding, to
asynchronous, non-slotted RA have also been proposed and
analyzed~\cite{Akyildiz2021,Clazzer2022}. Further improvement is
obtained by exploiting the capture effect in wireless channels, which
can be achieved by introducing transmit power
diversity~\cite{Zhao2020} or exploiting the natural fading, as
in~\cite{Mengali2017,Clazzer2018,Alvi2018}, or exploiting power-domain
non-orthogonal signaling techniques~\cite{Shao2019}. Another form
increasing the system capacity and EE of coded ALOHA is to improve the
capability of the SIC-based decoder. Instead of decoding a single
packet per slot, SIC can be used to attempt the recovery of up to $k$
packets simultaneously on the same slot, by using techniques like
those in power-domain non-orthogonal multiple access~\cite{Liu2022},
or a pure signal-based processing like in~\cite{Kazemi2022} that
avoids decoding the packet symbols before the cancellation of
interference. Another possibility is to equip the receiver with an
user-activity detection algorithm, since multiple-packet reception is
conceptually very similar to grant-free access
schemes~\cite{Srivatsa2021}. This $k$-MUD (multiuser detection)
capability can therefore increase the probability of decoding
successfully the packets, reduces the delay and enlarges the capacity
of the system (up to a factor $k$, ideally) since several packets can
be decoded at each round. However, the complexity of a $k$-MUD
receiver is higher.

Performance analysis of IRSA and other coded ALOHA RA protocols
exploits the conceptual connection between the SIC process and
generalized iterative decoding of graph-based codes (e.g., low density
parity-check codes
(LDPC))~\cite{Luby2001,Richardson2001,Di2002,Paolini2015}. There, the
evolution over time of the probabilities of correct decoding of a
received codeword is derived. It runs out to be a particular 
form of a message-passing algorithm over a bipartirte graph. This density 
evolution (DE) forms a system of discrete dynamical equations whose
fixed-point solutions yield the asymptotic error probability of the
decoding process. In IRSA, a completely analogous approach can be
followed for its performance analysis~\cite{Liva2011,Sandgren2017},
using the same tools as in graph-based coding theory, and a similar
pattern of performance is obtained: the packet loss ratio (decoding
failures at the SIC-receiver) follows two regimes, a waterfall region
of fast, exponential decay, and an error-floor region where
performance improvement practically stops. The latter is due to the
existence of stopping sets in the underlying graph, i.e., the receiver
reaches a trapping state where SIC cannot further proceed with the
cancellation of interference. Since the analysis of the DE is complex,
other works focus on the asymptotic performance exclusively, for a
number of scenarios~\cite{Liva2011}, or have explored alternative
tools for the derivation of the asymptotic performance. A more general
approach for analyzing coded RA in the asymptotic regime is the
Poisson receivers framework~\cite{Yu2021,Chang2021a}, which is able to
capture spatial and temporal coupling among a set of receivers,
including coding. However, the modeling and analysis of the DE is
still of interest for deriving good approximations to finite-length
systems and for gaining better insight into the design of coded ALOHA
protocols, especially when the receiver uses more sophisticated
strategies for SIC.

In this paper, we introduce a DE analysis of $k$-MUD IRSA that allows a
full characterization of performance of these coded MA systems. Our 
approach contributes to the state-of-the-art as follows:
\begin{itemize}
\item We apply known tools for the derivation of the DE for $k$-MUD IRSA, 
both for the case of uniform choice of the slots for transmission of the replicas, 
and for the case in which the devices transmit the first replica immediately after 
the packet generation.
  
\item Our results extend and improve the results in~\cite{Ghanbarinejad2013,Hmedoush2020,Dumas2021}, 
and also provide a rigorous support to some of the examples discussed under the Poisson 
receivers model in~\cite{Yu2021}. These works deal mainly with the asymptotic regime and consider 
its stability~\cite{Yu2021} and the optimization of the degree distributions~\cite{Hmedoush2020}. 
The approach in~\cite{Liva2011} relies on the classical Poisson approximation, i.e., a 
binomial distribution can be approximated by a properly parameterized Poisson distribution. 
\cite{Dumas2021} derives the density evolution only for the $2$-MUD case. This work makes an 
exact derivation of the DE for any $k$.
  
\item Unlike previous works in the literature, where the analysis focused on an asymptotically
large number of slots (e.g.,~\cite{Baiocchi2018,Hmedoush2020,Stefanovic2018}), our results hold 
for any finite frame length $n$. This entails that the asymptotic behavior of the previously mentioned 
variants (immediate transmission in the first slot and randomized selection of the first slot) 
have the same performance, as expected. 
  
\item We provide new bounds to the system capacity of $k$-MUD IRSA. While it is 
known~\cite{Hmedoush2020} that $\mathsf{G} / k < 1$, where $\mathsf{G}$ is the offered traffic, 
we improve the known bounds.

\item We show via numerical experiments that $k$-MUD can attain significantly better packet 
loss ratio and lower delay than IRSA for low values of $k$, without having to optimize with 
complex design or search the degree distributions.
\end{itemize}

The rest of the paper is organized as follows. The system model is
introduced in Section~\ref{sec:model}. In
Section~\ref{sec:density-evolution}, we present the results on the
density evolution of IRSA with MUD
capability. Section~\ref{sec:bounds} discusses the bounds on the
system capacity based on our DE analysis, and
Section~\ref{sec:results} provides numerical performance results. The
conclusions are summarized in Section~\ref{sec:conclusions}.

\section{System Model}
\label{sec:model}

\subsection{Irregular repetition slotted Aloha}
\label{sec:IRSA}

We consider an irregular repetition Slotted Aloha (IRSA) system with a
variable number of users transmitting to a common receiver, the access
point (AP). In IRSA, a degree-$\mathsf{r}_u$ user intending to
transmit a new packet selects a repetition factor $\mathsf{r}_u$
probabilistically, according to a predefined distribution represented
by the polynomial moment generating function
\begin{equation}
  \Lambda(x) = \sum_{i = 1}^\mathsf{d} \Lambda_i x^i,
\end{equation}
where $\Lambda_i$ is the probability of transmitting $i$ replicas, and
$\mathsf{d}$ is the maximum number of copies for a packet, with
$\sum_{i = 1}^\mathsf{d} \Lambda_i = 1$. The average number of copies
per packet is therefore
$\Esp[{\mathsf{D}}] = \Lambda^\prime(1) = \sum_{i = 1}^\mathsf{d} i
\Lambda_i$, where $\Lambda^\prime(\cdot)$ is the derivative of
$\Lambda(x)$.  We assume that the new packets have a transmission
time equal to one slot and are generated according to a Poisson
process of intensity $\lambda_a$ packets per slot. Thus, the number
$\mathsf{N}$ of new packets per unit time follows a probability mass
function
\begin{equation}
  \Prob(\mathsf{N} = j) = e^{-\lambda_a} \frac{\lambda_a}{j!}, \qquad
  j = 0, 1, \dots 
\end{equation}

\begin{figure}[t]
  \centering
  \includegraphics[width=\textwidth]{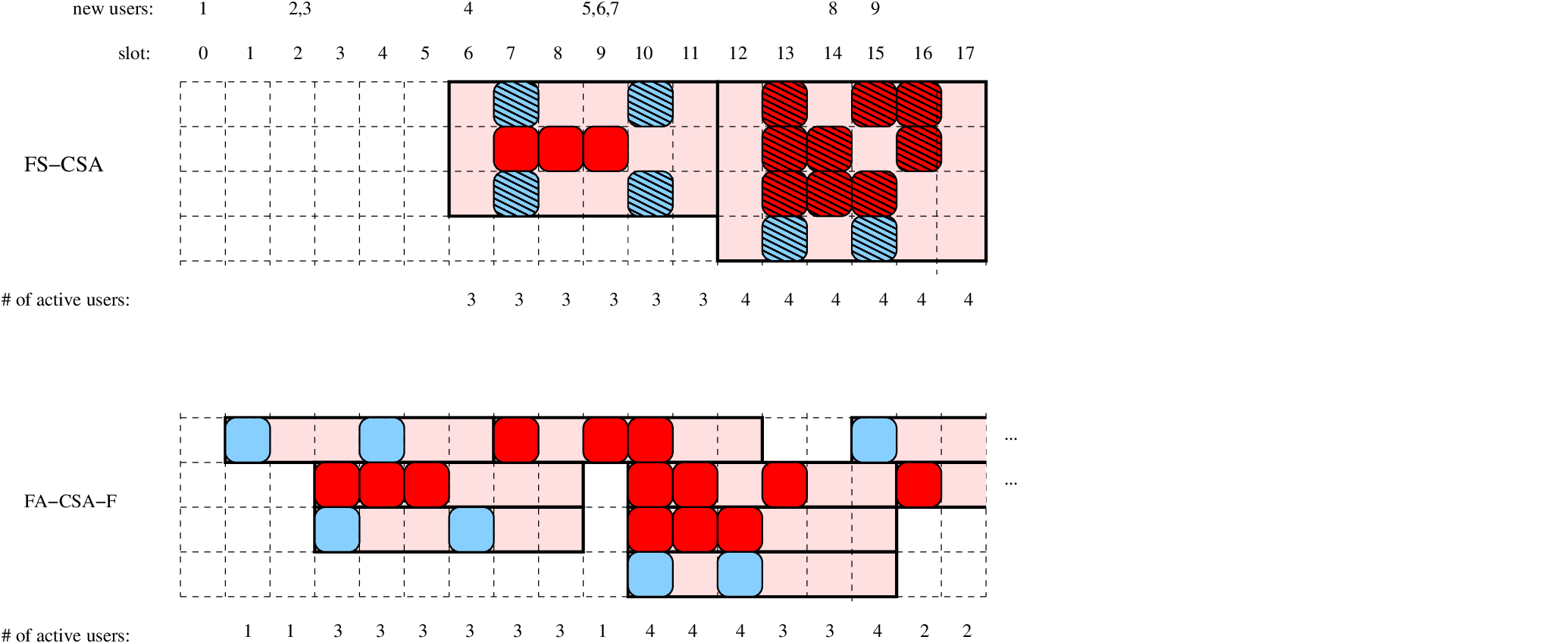}
  \caption{\label{fig:IRSA-example1} An example of frame asynchronous
    IRSA for $n = 6$. The asynchronous system fixes the first packet
    copy at the first slot.}
\end{figure}

We analyze in this paper two variants of the frame asynchronous (or
frameless)~\cite{Akyildiz2021} IRSA. In both cases, all users'
transmissions are arranged into frames consisting of $n$ slots each. A
degree-$\mathsf{r}_u$ user who joins the system at time $t$ places its
transmissions within the subsequent $n$ slots after the packet
arrival, but an active user either transmits the first replica of its
new packet in the next slot, and distributes the remaining
$\mathsf{r}_u - 1$ copies randomly across the $n - 1$ subsequent slots
or, alternatively, distributed its $\mathsf{r}_u$ replicas uniformly
over the user's virtual frame $[t + 1, t + n]$, without enforcing a
transmission immediately in the first slot. In all the cases, the
duration $n$ of the frames is a system-wide parameter common to all
users and known to the receiver. Fig.~\ref{fig:IRSA-example1} shows an
example of activity in this asynchronous operation mode.

\subsection{SIC-based Decoding}
\label{sec:decoding}

We assume the receiver has multi-user detection (MUD) capability of
degree $k$ and moreover that it adopts a successive interference
cancellation (SIC) strategy for decoding. More specifically, it is
assumed that, when a slot contains $k$ or less user packets all of
these can be recovered by the AP perfectly, but that decoding will
fail whenever the slot has more than $k$ active users. Note that the
classical destructive collision channel model is the case $k = 1$,
whereas a $k$-MUD receiver tolerates a certain level of interference
from other users. The assumption that $k$ is fixed also reflects well
the case where all users transmit with fixed power, or when there is
some form of power control enforced by the AP so that the received
signals from different users have approximately equal power. The
throughput and stability properties of slotted Aloha with MUD have
been studied in many preceding works, e.g.~\cite{Ghez1988}, and an
efficient form of using SIC for collision resolution in the delay
domain, a natural choice for SA, is developed in~\cite{Kazemi2022}.

If the receiver decodes successfully the packets in a slot, then it
iteratively selects any other degree-$k$ or less slot in its memory,
and decodes the packets in that slot by cancelling the interference of
the previously decoded packets. This process continues until no new
degree-$k$ or lower slots are found or until a maximum 
number of iterations is reached (to limit the memory and delay). The only 
difference between the synchronous and the asynchronous modes of operation 
is that, in the latter, the AP needs to track all the slots in the entire 
history of the system. So, in order to limit memory and decoding delay, a
finite sliding-window of $\mathsf{W}$ slots is usually employed in the
decoder.

\begin{figure}[t]
  \centering
  \includegraphics[width=\textwidth]{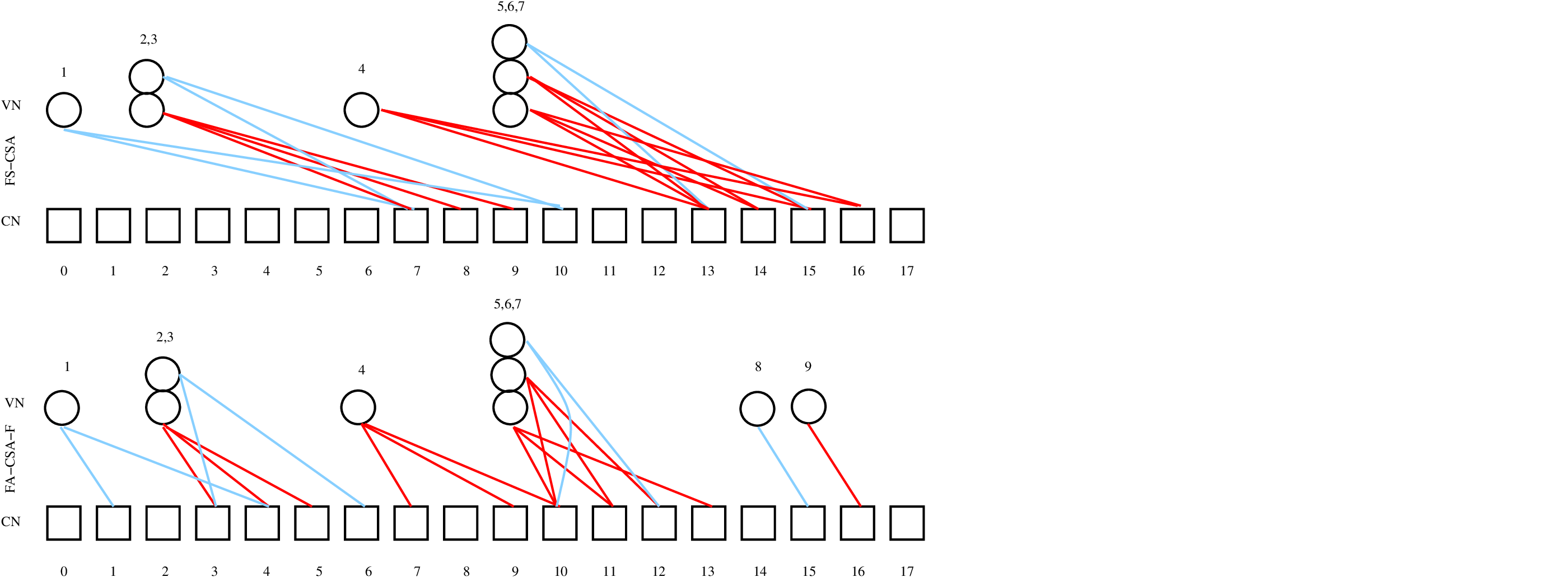}
  \caption{\label{fig:SIC-decoding} Equivalent graph representation of
    the system depicted in Figure~\ref{fig:IRSA-example2}.}
\end{figure}

\begin{figure}[t]
  \centering
  \includegraphics[width=\textwidth]{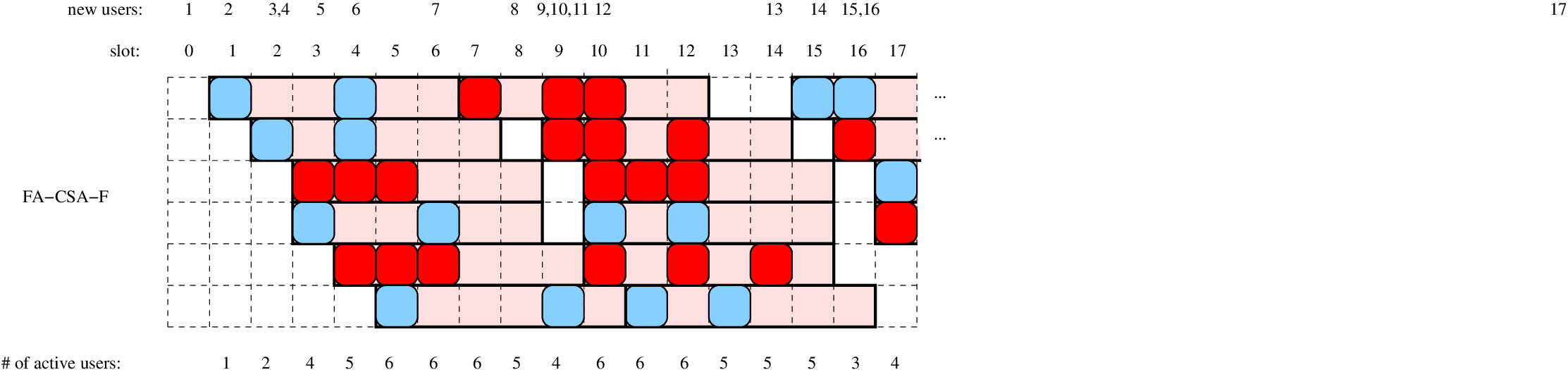}
  \caption{\label{fig:IRSA-example2} An example of asynchronous IRSA
    with $k = 2$.}
\end{figure}

The SIC-based decoding process in IRSA is entirely analogous to a
message passing algorithm in a factor graph~\cite{Feng2022}. Hence, it
may be modeled by a bipartite graph
$\mathcal{G} = \{ \mathcal{V} \cup \mathcal{C}, \mathcal{E} \}$
analogous to iterative decoding for graph-based
codes~\cite{Richardson2001,Ivanov2017} (e.g., LDPC codes), where
$\mathcal{V}$ is the set of variable nodes in the factor graph,
$\mathcal{C}$ is the set of check or constraint nodes and
$\mathcal{E}$ are the edges connecting the variable and check
nodes. For IRSA decoding, edge $(i, j) \in \mathcal{E}$ represents
user $i$ transmitting one of its replica packets in slot $j$.
Fig.~\ref{fig:SIC-decoding} depicts the graph model corresponding to
the example in Fig.~\ref{fig:IRSA-example1}. In the graph, the colors
just differentiate between users with replica factors
$\mathsf{r}_u = 2$ or $\mathsf{r}_u = 3$. An example in which
SIC-based decoding succeeds in recovering all the packets if
$k \geq 2$ is shown in Fig.~\ref{fig:IRSA-example2}.


For the analysis of the density evolution on the underlying system
graph, it is customary to introduce the enumerator polynomials
\begin{align*}
  \Lambda(x) \triangleq \sum_{i=1}^\mathsf{d} \Lambda_i x^i, &\quad
  \Gamma(x) \triangleq \sum_{j = 0}^{\mathsf{N}_a} \Gamma_j x^j \\
  \lambda(x) \triangleq \sum_{i = 1}^\mathsf{d} \lambda_i x^{i-1},
  &\quad \gamma(x) \triangleq \sum_{j = 0}^{\mathsf{N}_a} \gamma_j x^{j - 1}
\end{align*}
where $\Gamma_j$ is the probability that a decoding node has degree
$j = 0, \dots, \mathsf{N}_a$; and $\lambda(x), \gamma(x)$ are the
weight enumerators of the degree distributions of user and decoding
nodes, respectively, from an edge perspective. Namely, $\lambda_i$
denotes the probability that a random edge is connected to a degree-$i$
variable node, and similarly $\gamma_j$ is the probability that the random
edge is connected to a degree-$j$ check node. These probabilities are
given by
\begin{equation}
  \lambda_i = \frac{i \Lambda_i}{\sum_m m \Lambda_m}, \quad \gamma_j = \frac{j
    \Gamma_j}{\sum_m m \Gamma_m}.
\end{equation}
Hence, $\lambda(x) = \Lambda^\prime(x) / \Lambda^\prime(1)$ and
$\gamma(x) = \Gamma^\prime(x) / \Gamma^\prime(1)$.

The fundamental performance measures for this system are the following:
\begin{itemize}
\item The packet loss ratio $\mathsf{PLR}$ is the
  probability that the packet of an arbitrary users is never recovered
  by the SIC-based decoder.

\item The delay $\mathsf{T}$ of a resolved packet is the number of slots between
  the user's arrival and the slot where the packet is resolved and
  correctly decoded.
\end{itemize}

\section{Analysis of the density evolution}
\label{sec:density-evolution}

\subsection{Density evolution}

In IRSA, packet losses arise from activity patterns that the SIC-based
decoder is unable to resolve. As explained above, with slotted ALOHA
these patterns are equivalent to stopping sets of LDPC
codes~\cite{Richardson2001,Di2002}, so they can be analyzed with tools
borrowed from the theory of codes on graphs, in particular with the
density evolution (DE) technique. The DE is basically a set of
fixed-point equations that describe the change over time of the
probabilities $p_{(i,j)}$ ($q_{i,j}$) of the messages exchanged from a
variable node (check node) to a check node (variable node) over the
edge $(i, j)$ in the factor graph.  While the exact evaluation of the
DE is not strictly necessary for computing the $\mathsf{PLR}$ in more
general settings than MUD~\cite{Yu2021} (including correlated
receivers or transmitters), a closed formula of the DE for a $k$-MUD
receiver can be obtained and gives further insights on the system
performance in the asymptotic regime $n \to \infty$. Moreover,
accurate approximations to the $\mathsf{PLR}$ can be derived from the
asymptotic results once the DE is known~\cite{Sandgren2017,Amat2018}.

We start recalling from~\cite{Sandgren2017} the degree distribution of
a check node at epoch $i \in \mathbb{Z}^+$, i.e., the probability
distribution of the number of received packets seen by the AP in slot
$i$.
\begin{theorem}[\cite{Sandgren2017}, Propositions 1 and 2]
  \label{th:degree-dist}
  The degree distribution for a decoding node at slot $i = 1, 2, \dots$
  is given by:
  \begin{enumerate}
  \item[(i)] For frame asynchronous IRSA with transmission of the
    first replica at the initial slot
    \begin{align}
      \Gamma^{(i)}(x) &=\gamma^{(i)}(x) = \exp \bigl( -\lambda_a (1
      - x) \bigr), \quad i = k n
      \label{eq:cn-dist-a} \\
      \Gamma^{(j)} (x) &= \rho^{(j)}(x) = \exp \left( -
        \frac{ \delta_j \bigl( \Lambda^\prime(1) - 1 \bigr)}{n - 1} (1
        - x) \right), \quad j \neq kn \label{eq:cn-dist-b}
    \end{align}
    where
    \begin{equation}
      \label{eq:delta-i}
      \delta_j = \begin{cases}
        \lambda_a \min \{ j - 1, n - 1 \}, & \quad\text{with start-up phase} \\
        \lambda_a (n - 1), & \quad\text{without start-up phase},
      \end{cases}
    \end{equation}
    where~\eqref{eq:cn-dist-a} corresponds to the first slot in a
    frame, and~\eqref{eq:cn-dist-b} is for the subsequent
    slots in the frame.

  \item[(ii)] For frame asynchronous IRSA with uniformly distributed replicas
    \begin{equation}
      \Gamma^{(i)}(x) = \rho^{(i)}(x) = \exp \biggl( -\frac{\mu_i}{n}
      \Gamma^\prime(1) (1 - x) \biggr)
    \end{equation}
    where
    \begin{equation}
      \mu_i = \begin{cases}
        i \lambda_a, &\quad \text{for $1 \leq i < n$} \\
        n \lambda_a, &\quad \text{for $i \geq n$}
      \end{cases}
    \end{equation}
    with start-up phase, and $\mu_i = n \lambda_a$ without start-up
    phase.
  \end{enumerate}
\end{theorem}
The term start-up phase refers to a system that starts totally empty,
i.e., without active users and with no packets waiting to be
decoded. Conversely, a system without start-up phase is supposed to be
already in steady state, and analyzed thereafter. Observe from
Theorem~\ref{th:degree-dist} that the effect of the start-up phase
vanishes after $n$ slots (one frame), so it does not have an impact on
the $\mathsf{PLR}$. Also, \eqref{eq:cn-dist-b} differs
from~\eqref{eq:cn-dist-a} in that the factor
$\Lambda^\prime(1) - 1 = \mathbb{E}[\mathsf{D}] - 1$ is absent for the
start-phase.

Our main result is the derivation of the density evolution of the
irregular repetition slotted ALOHA with multipacket reception.
\begin{theorem}
  \label{thm:de}
  For any $i, j \geq 1$, denote by $p_{(i, i)}$, $p_{(i, j)}$,the
  probabilities of decoding failure of a packet replica along edges
  $(i, i)$, $(i, j)$ at a user node. And let $q_{(i, i)}$ and $q_{(i, j)}$
  be the decoding failure probabilities on edges $(i, i)$, $(i, j)$ at
  receiver node $i$.  The density evolution equations for the $k$-MUD
  IRSA with transmission in the first slot are given by:
  \begin{align}
    q_{(i, i)} = q_{(i, j)} &= 1 - \sum_{m = 0}^{k - 1} \exp(-\nu_i)
    \frac{\nu_i^m}{m!}
    \label{de-irsa-fs-a} \\
    p_{(i, i)} &= \Gamma^{(i)} \bigl(\overline{q}_{(i,
      i)} \bigr) \label{de-irsa-fs-b} \\
    p_{(i, j)} &= q_{(i, i)} \gamma^{(i)} \bigl(\overline{q}_{(i,
      i)} \bigr) \label{de-irsa-sa-c} \\
    \overline{q}_{(i, i)} &= \frac{1}{n - 1} \sum_{j = i + 1}^{i + n
      - 1} q_{(j, i)} \label{de-irsa-fs-d}
  \end{align}
  where $\nu_i = \lambda_a p_{(i, i)} + \frac{\Lambda^\prime(1)}{n -
    1} \overline{p}_i \delta_i$ and $\overline{p}_i$ is
  \begin{equation}
    \begin{cases}
      \sum_{j = 1}^i \frac{p_j}{i}, \quad & 1 \leq i < n \\
      \sum_{j = i - n + 1}^i \frac{p_j}{n}, \quad & i \geq n.
    \end{cases}
  \end{equation}
  The density evolution equations for the $k$-MUD IRSA with uniformly
  distributed replicas is given by
  \begin{align}
    q_i &= 1 - \sum_{m = 0}^{k - 1} \exp(-\eta_i) \frac{\eta_i^m}{m!}
    \label{de-irsa-a} \\
    p_i &= \lambda(\overline{q}_i) \label{de-irsa-b} \\
    \overline{q}_i &= \frac{1}{n} \sum_{j = i}^{i + n - 1}
    q_j \label{de-irsa-c} 
  \end{align}
  where $\eta_i = \mu_i \overline{p}_i \Lambda^\prime(1) / n$ and
  $\overline{p}_i = (\sum_{j = i - n + 1}^i p_i) / n$ for $i \geq n$
  or $\overline{p}_i = (\sum_{j = 1}^i p_j) / i$ if $1 \leq i < n$.
\end{theorem}
\begin{proof}
  See~\ref{app:proof-a}.
\end{proof}
The special case $k = 1$ gives the results of pure IRSA
(packet-by-packet SIC decoding) obtained previously
in~\cite{Sandgren2017}. Note that the simultaneous decoding of $k$
packets changes only the probability of decoding failure at the
receiver nodes through~\eqref{de-irsa-fs-a} and~\eqref{de-irsa-a}.

\begin{remark}
  The density evolution only displays minor differences between the
  uniformly temporally spread replicas and the use of the first
  slot. In functional form, \eqref{de-irsa-fs-a} and~\eqref{de-irsa-a}
  are equal, except for the respective arrival
  rates. Equations~\eqref{de-irsa-fs-d} and~\eqref{de-irsa-c} are
  obviously identical, and the only changes
  between~\eqref{de-irsa-b}-\eqref{de-irsa-c} and~\eqref{de-irsa-b}
  arise from the deterministic use of the first
  slot (cf.~\eqref{de-irsa-fs-b}) and also from the fact that the
  degree distribution as seen from the user node has to be modified to
  account for this first transmission ($\Lambda^{(i)}$ and
  $\lambda^{(i)}$, respectively). Note also that the averaging of
  probabilities in~\eqref{de-irsa-fs-d} and~\eqref{de-irsa-c} spans a
  window of $n$ previous slots, where $n$ is the frame length.
\end{remark}
\begin{remark}
  Theorem~\ref{thm:de} gives the probabilities $p_i, q_i$ from the
  node and slot perspectives \emph{for any value of} $i$, not only
  when $i \to \infty$. The explicit dependence on the probabilities of
  the $n$ previous nodes or slots is captured via the definition of
  the averages $\overline{p}_i$ and $\overline{q}_i$. These averages
  obviously involve the same number of terms after the first frame,
  i.e., for $i > n$.
\end{remark}

Observe that the numerical calculation for the DE in
Theorem~\ref{thm:de} involves only polynomial evaluations at each
time.  In the asymptotic regime $i \to \infty$, we obtain the
$\mathsf{PLR}$ of $k$-MUD IRSA as an immediate consequence of the
latter Theorem.
\begin{corollary}
  The packet loss ratio of $k$-MUD IRSA at receiver node $i$ is
  \begin{enumerate}
  \item[(i)] With first slot fixed
    \begin{equation}
      \overline{p}_i = \Lambda(\overline{q}_{i}) q_{(i, i)} /
      \overline{q}_i;
    \end{equation}

  \item[(ii)] and with uniformly distributed replicas
    \begin{equation}
      \overline{p}_i = \Lambda(\overline{q}_i).
    \end{equation}
  \end{enumerate}
  The asymptotic $\mathsf{PLR}$ is
  $\mathsf{PLR} = \lim_{i \to \infty} \overline{p}_i$.
\end{corollary}
In practice, the $\mathsf{PLR}$ can be computed as the fixed-point
solution of~\eqref{de-irsa-fs-a}-\eqref{de-irsa-fs-d}
or~\eqref{de-irsa-a}-\eqref{de-irsa-c}, respectively, setting
initially all the edge probabilities to $1$.

\subsection{Potential function and stability}

For $i \gg n$, the density evolution for asynchronous IRSA converges
to unique values $p$, $q$, independent of the slot position $i$, which
are the solution to the fixed point equation
\begin{equation}
  x = \Lambda \bigl( g_k(x) \bigr)
\end{equation}
where
\begin{equation}
  \label{eq:potential-g}
  g_k(x) = 1 - \exp(-\zeta x) \sum_{j = 0}^{k - 1} \frac{(\zeta x)^j}{j!},
\end{equation}
$\zeta\triangleq \lambda_a \Lambda^\prime(1)$, and $\lambda_a$ is
the arrival rate of transmitters per slot. Recall from
Theorem~\ref{thm:de} that
\begin{equation}
  \label{eq:fixed-point}
  p_i = \Lambda \biggl( 1 - \exp(-\eta_i) \sum_{m = 0}^{k - 1}
  \frac{\eta_i^m}{m!} \biggr)
\end{equation}
and $\eta_i = \lambda_a \overline{p}_i (\Lambda^\prime(1) - 1)$, where
$\overline{p}_i$ is the average of $\{ p_i \}$ over the last $n$
slots. The mapping $\eta_i \to T(\eta_i) := \exp(-\eta_i) \sum_{m =0}^{k - 1}
\eta_i^m / m!$ is continuous and increasing, and has $T(0) =
1$. Assume now that $p_{i + 1} \leq p_i$ for all $i < s$. Then
$\eta_{s+1} \leq \eta_s$ since the average $\overline{p}_i$ is
non-increasing, and
\begin{equation}
  p_{s+1} = \lambda\bigl( 1 - T(\eta_s) \bigr) \leq \lambda\bigl( 1 -
  T(\eta_{s-1}) \bigr) = p_s
\end{equation}
where the inequality follows since $\lambda(x)$ is increasing in
$x \in [0, 1]$, $\eta_s \leq \eta_{s+1}$ and the assumption that
$T(\cdot)$ is continuous and increasing. Therefore, the sequence $\{ p_s \}$ and its average
$\overline{p}_s$ converge to a limit $p^\ast \geq 0$ which is
necessarily a fixed point of the DE.

In the context of graph-based codes analyzed through the DE technique,
the potential function is defined as~\cite{Richardson2001}
\begin{equation}
  U(x) := x g(x) - G(x) - F\bigl(g(x) \bigr)
\end{equation}
where $G(x) := \int_0^x g(z) dz$ and $F(x) = \int_0^x \lambda(z)
dz$. Recalling that
$\lambda(x) = \Lambda^\prime(x) / \Lambda^\prime(1)$, we have
$F(x) = (\Lambda(x) - \Lambda(0)) / \Lambda^\prime(1) = \Lambda(x) /
\Lambda^\prime(1)$.  The usefulness of this function for the stability
of the system comes from the fact that
\begin{equation}
  \label{eq:threshold}
  \lambda^\star_a = \sup \left( \lambda_a: U^\prime(x) > 0 \forall x \in [0, 1] \right)
\end{equation}
is the threshold for the feasible arrival rates, i.e, the supremum of
the arrival rates such that
$\mathsf{PLR} = \lim_{i \to \infty} p_i = 0$.\footnote{Recall that the
  $\mathsf{PLR}$ accounts for the decoding errors with SIC,
  exclusively. We do not include in the model the possibility of
  transmission errors for simplicity.} Note that for computing
$\lambda^\star_a$ we can use
$\tilde{F}(x) = \Lambda(x) / \Lambda^\prime(1)$ in the definition of
the potential function. The closed-form expression for the potential
function in $k$-MUD IRSA is given in the following theorem.
\begin{theorem}
  \label{thm:potential-function}
  The potential function for $k$-MUD frame asynchronous IRSA with is
  given by
  \begin{equation}
    U_k(x) = k - \frac{1}{\zeta} \exp(-\zeta x) \sum_{j = 0}^{k - 1}
    \frac{(\zeta x)^j}{j!} (\zeta x + k - j) - \frac{\Lambda\bigl( g_k(x)
      \bigr)}{\Lambda^\prime(1)}
  \end{equation}
  where $\zeta \triangleq \lambda_a \Lambda^\prime(1)$and $g_k(x)$ is
  given in~\eqref{eq:potential-g}.
\end{theorem}
\begin{proof}
  First, consider the function $g_k(x)$ in~\eqref{eq:potential-g}.
  Integrating by parts each of the terms in the sum we get
  \begin{multline*}
    I_j(x) \triangleq \int_0^x \exp(-\zeta t) \frac{(\zeta t)^j}{j!} dt =
    I_{j-1}(x) - \frac{1}{\zeta} \exp(-\zeta x) \frac{(\zeta x)^j}{j!} = \dots = I_0(x) - \frac{1}{\zeta} \exp(-\zeta x) \sum_{m = 1}^j
    \frac{(\zeta x)^m}{m!}
  \end{multline*}
  Using that $I_0(x) = 1 - 1/\zeta \exp(-\zeta x)$ we obtain
  \begin{equation*}
    I_j(x) = 1 - \frac{1}{\zeta} \exp(-\zeta x) \sum_{m = 0}^{j}
    \frac{(\zeta x)^j}{j!} 
  \end{equation*}
  and thus
  \begin{multline*}
    G(x) = x - \sum_{j = 0}^{k - 1} I_j(x) = x - k + \frac{1}{\zeta}
    \exp(-\zeta x) \sum_{j = 0}^{k - 1} \sum_{m = 0}^j \frac{(\zeta x)^m}{m!} = x - k
    + \frac{1}{\zeta} \exp(-\zeta x) \sum_{j = 0}^{k - 1} (k - j)
    \frac{(\zeta x)^j}{j!}, 
  \end{multline*}
  where the last equality follows after exchanging the order of the
  summations. Now, since
  \begin{equation*}
    x g_k(x) = x - \frac{1}{\zeta} \exp(-\zeta x) \sum_{j = 0}^{k - 1}
    \frac{(\zeta x)^{j + 1}}{j!}
  \end{equation*}
  we have
  \begin{equation*}
    U_k(x) = xg_k(x) - -G(x) - F\bigl( g_k(x) \bigr) = k - \frac{1}{\zeta} \exp(-\zeta x) \sum_{j = 0}^{k - 1} \frac{(\zeta x)^j}{j!} (\zeta x + k - j) - \frac{\Lambda\bigl( g_k(x) \bigr)}{\Lambda^\prime(1)}.
\end{equation*}

  This finishes the proof.
\end{proof}
The special case $k = 1$ yields the formula
\begin{equation}
  U_1(x) = 1 - \frac{\zeta x + 1}{\zeta} \exp(-\zeta x) -
  \frac{\Lambda\bigl( g_1(x) \bigr)}{\Lambda^\prime(1)}.
\end{equation}
\begin{figure}[t]
  \centering
  \begin{subfigure}{0.49\textwidth}
    \includegraphics[width=9cm]{%
      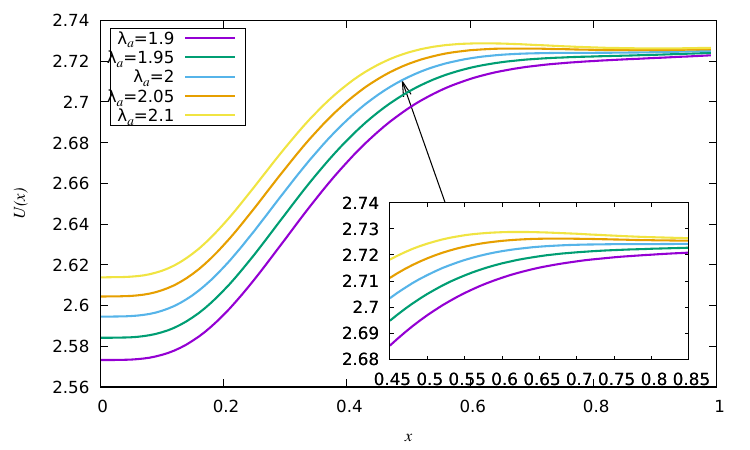}
    \caption{ $\Lambda_1(x) = 0.86 x^3 + 0.14 x^8$.}
  \end{subfigure}
  \begin{subfigure}{0.49\textwidth}
    \includegraphics[width=9cm]{%
      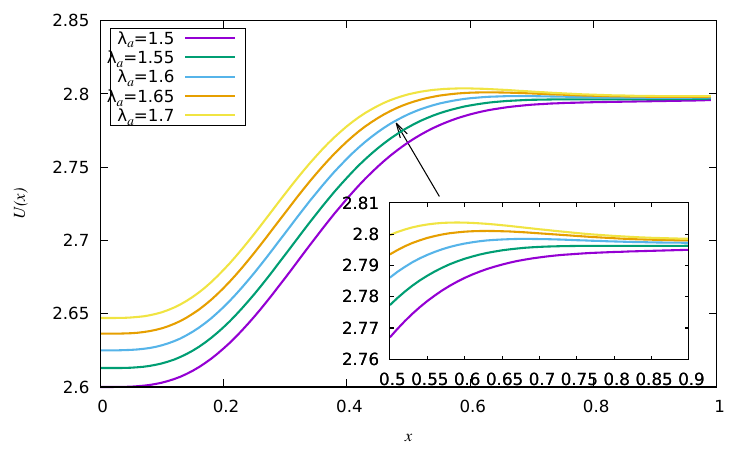}
        \caption{$\Lambda_0(x) = x^5$.}
  \end{subfigure}
  \caption{\label{fig_potential} Examples of the potential function
    for FA-IRSA with uniform slot selection vs. users arrival rate
    $\lambda_a$; $k = 3$.}
\end{figure}
In Figure~\ref{fig_potential} we show an example of the potential
function for the $k$-MUD IRSA with uniform slot selection, for two
degree distributions $\Lambda(x) = 0.86 x^3 + 0.14 x^8$ and
$\Lambda(x) = x^5$, and $k = 3$, as a function of the arrival rate
$\lambda_a$. Visually, the threshold $\lambda^\ast \approx 2$ for the
first case, and is approximately $1.65$ for the latter case.

\section{Capacity bounds}
\label{sec:bounds}

Theorem~\ref{thm:potential-function} provides a closed-form expression
for the potential function. In principle, one could use it for
determining analytically the maximum arrival rate $\lambda_a^\ast$ to
the system, simply by finding whether the maximum of $U_k(x)$ is
attained in $(0, 1)$ or not. Using~\eqref{eq:threshold}, one can
conclude that the arrival rate $\lambda_a$ is feasible for decoding
without error if $\lambda(g_k(x)) < x$ for all $x \in (0, 1)$. In
fact, a direct calculation from Theorem~\ref{thm:potential-function}
arrives at the same conclusion.
\begin{lemma}
  \label{lemma:maximum}
  The maximum of the potential function $U_k(x)$ is the solution to
  \begin{equation*}
    \lambda \bigl( g_k(x) \bigr) = x.
  \end{equation*}
\end{lemma}
\begin{proof}
  See~\ref{app:proof-b}.
\end{proof}
This simple result can be used in two ways. First, we can use the
expression in the Lemma by expanding the term $\lambda(g_k(x))$ as a
power series in the variable $x$ and equate the terms in the RHS and
the LHS so that the LHS is less than $\lambda_a^{-1}$ for any
$x \in (0, 1)$. This would imply that the potential function is
increasing in the interval, hence the arrival rate $\lambda_a$ is
feasible for some degree distribution $\Lambda(x)$ which is the
solution to a set of \emph{linear} equations. In other words, it is
possible to use the result of the Lemma to synthesize optimal degree
distributions just by solving numerically a few linear equations under
the constraints that $\sum_i \lambda_i = 1$ and $\lambda_i \geq
0$. This improves the method described in~\cite{Hmedoush2020}, where a
linear program has to be solved for the same purpose. We can further
note that the optimal distributions are of the form
$\lambda(x) = o(x)$ (see~\cite{Hmedoush2020, Sandgren2017}). This
implies that the solution to the functional equation
$\lambda(g_k(x)) = x$ is bounded away from zero, since otherwise, as
$g_k(x) = o(x^k)$, we would have $o(x^{k+1}) = o(1)$, which is not
possible for $k \geq 1$. A similar argument shows that the root is
also bounded away from $1$, since $\lambda(x)$ is a convex polynomial,
hence it is a superadditive function and
\begin{equation}
  \lambda\bigl( g_k(x) \bigr) + \lambda\bigl( 1 - g_k(x) \bigr) \leq
  \lambda\bigl( g_k(x) + 1 - g_k(x) \bigr) = \lambda(1) = 1,
\end{equation}
hence $\lambda(1 - g_k(x)) \leq 1 - \lambda(g_k(x))$.

In~\cite{Hmedoush2020}, it is shown that a lower bound for $\lambda_a - k$ is given by
\begin{equation}
    \mathsf{E} = \int_0^{s_k} (F_k^{-1}(x) - \alpha_k x) \mathrm{d}x,
\end{equation}
where $F_k(x) = 1 - \sum_{i = 0}^{k-1} x^i / i!$, and $\alpha_k$ is the root of $F_k^{-1}(x) - (F_k^{-1})^\prime(x) x = 0$. A direct calculation yields
\begin{equation}
 \mathsf{E} = \exp(-y) \sum_{i = 0}^{k - 1} \frac{(k - i) y^i}{i!} + \frac{y}{2} F_k(y)
\end{equation} 
where $\alpha_k = F_k(y)$. 

\section{Numerical Results}
\label{sec:results}

In this Section, we investigate the performance of $k$-MUD IRSA
through numerical experiments. Specifically, we consider the packet
loss ratio and the decoding delay for several illustrative degree
distributions of the replica process at the transmitters,
$\Lambda_0(x) = x^5$, $\Lambda_1(x) = 0.86 x^3+0.14x^8$ and 
$\Lambda_2(x) = 0.8793 x^2 + 0.003 x^7 + 0.1204 x^{11}$, and 
$\Lambda_3(x) = 0.929 x^2 + 0.07 x^{11}$. The first is a regular repetition pattern, 
since it generates exactly $5$ replicas per packet,
and this will allow us to test the impact of sparsity on the performance metrics; 
the second one  was used in~\cite{Ivanov2017}. It is not optimal, but provides a high 
threshold for a probability at the error floor around $10^{-5}$; $\Lambda_3(x)$ was 
derived  in~\cite{Hmedoush2020} numerically as the optimal distribution for 
$k = 2$; $\Lambda_4(x)$ was also found in~\cite{Hmedoush2020} as the optimal distribution 
for $k = 3$ with degree less than or equal to $11$.
Table~\ref{table:simulation-parameters} summarizes the
configuration parameters used in the tests. All our results are given
as a function of the normalized system load
$\mathsf{L} = \lambda_a / k \in [0, 1]$ to make the comparison among
different values of $k$ easier.

\begin{table}[t]
  \centering
  \caption{\label{table:simulation-parameters} Simulation parameters
    used for performance analysis.}
  \begin{tabular}{lc}
    \textsc{parameter} & \textsc{values} \\ \hline
    Degree distributions & $\Lambda_0(x) = x^5$ \\
                         & $\Lambda_1(x) = 0.86 x^3+0.14x^8$ \\
                         & $\Lambda_2(x) = 0.8793 x^2 + 0.003 x^7 + 0.1204 x^{11}$ \\ 
                         & $\Lambda_3(x) = 0.929 x^2 + 0.07 x^{11}$ \\
    $k$ & $\{ 1, 2, 3 \}$ \\
    $n$ (slots) &  $\{ 50, 100, 150, 200, 300, 400 \}$ \\ \hline
  \end{tabular}
\end{table}
\begin{figure}[t]
  \centering
  \begin{subfigure}{0.49\textwidth}
    \includegraphics[width=9cm]{%
      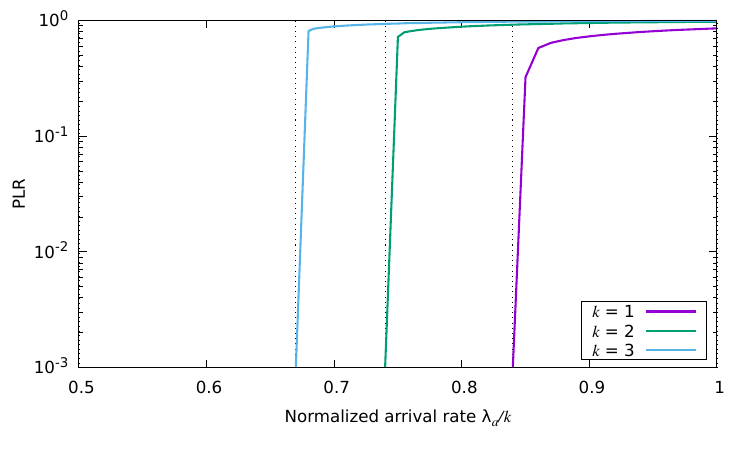}
    \caption{\label{fig:threshold_l_3_8_n_200:a} $\Lambda_1(x)$}
  \end{subfigure}
  \begin{subfigure}{0.49\textwidth}
    \includegraphics[width=9cm]{%
      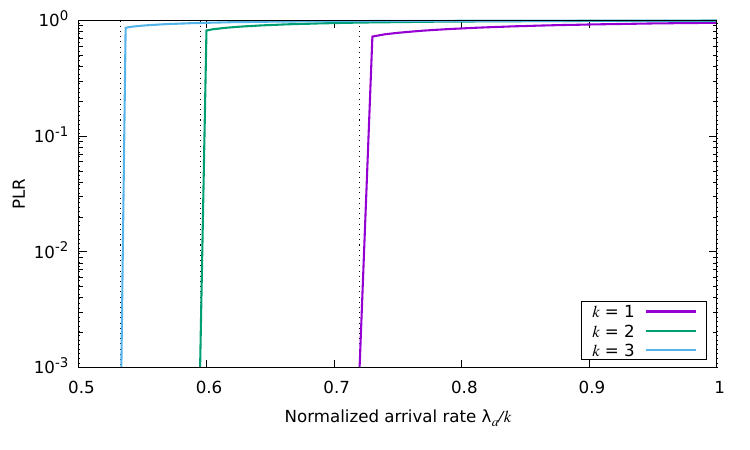}
    \caption{\label{fig:threshold_l_3_8_n_200:b} $\Lambda_0(x)$}
  \end{subfigure}
  \caption{\label{fig:threshold_l_3_8_n_200} Decoding thresholds for
    MUD IRSA, $k = 1, 2, 3$.}
\end{figure}

We start our numerical experiments by computing the asymptotic
iterative decoding thresholds separating the waterfall and error-floor
regions. These thresholds provide a simple numerical characterization
of the system stability. Figure~\ref{fig:threshold_l_3_8_n_200}
confirms the substantial benefits or irregular repetition (case (a))
over a constant factor for the replicas (case (b)), but also show that
the threshold $\lambda_a^\star$ is sublinear in $k$, suggesting that
the use of higher values of $k$ has a marginally decreasing
improvement. Thus, we limit our experiments to low values of $k$ in
the remaining tests, as listed in Table~\ref{table:simulation-parameters}.

\subsection{Packet Loss Rate and Average Delay: finite frame length}

\begin{figure}[t]
  \centering
  \begin{subfigure}{0.49\textwidth}
    \includegraphics[width=9cm]{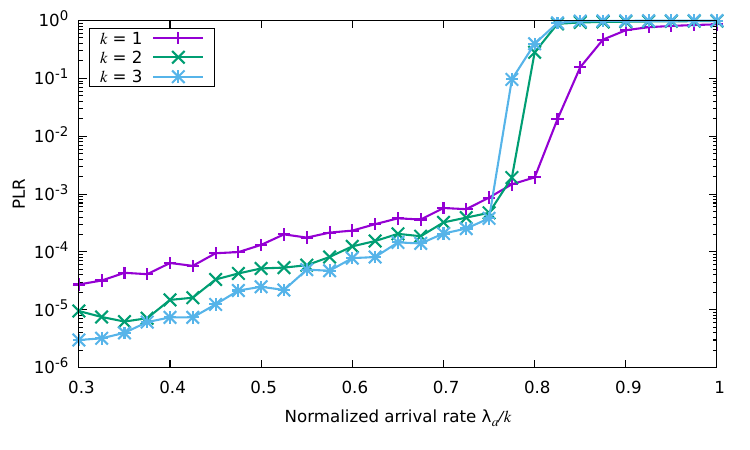}
    \caption{$\mathsf{PLR}$}
  \end{subfigure}
  \begin{subfigure}{0.49\textwidth}
    \includegraphics[width=9cm]{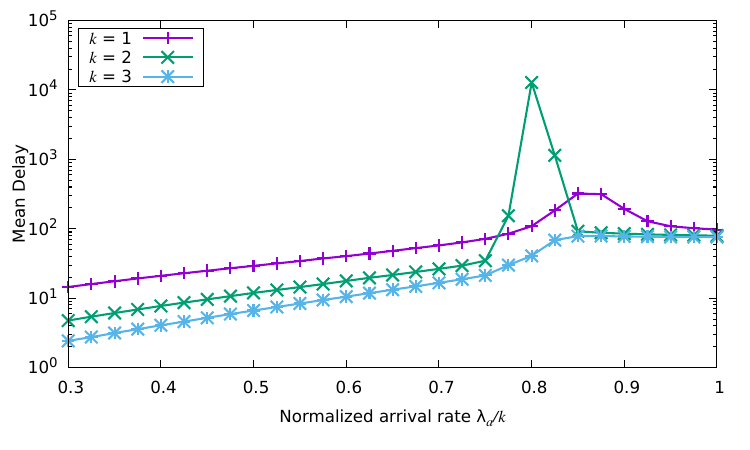}
    \caption{Average delay}
  \end{subfigure}
  \begin{subfigure}{0.49\textwidth}
    \includegraphics[width=9cm]{%
      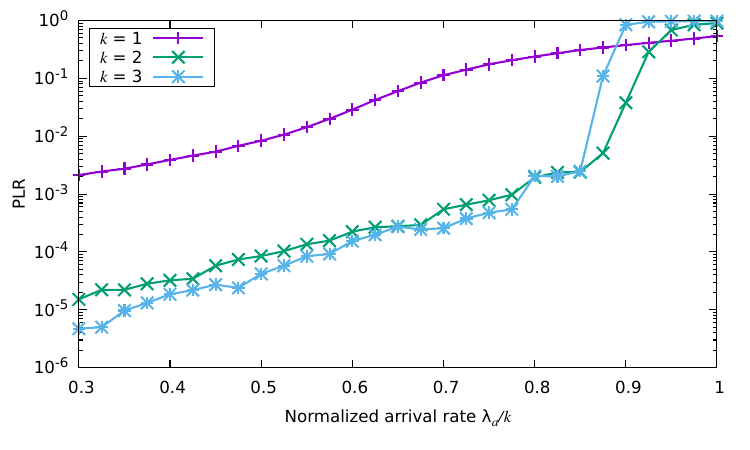}
    \caption{$\mathsf{PLR}$}
  \end{subfigure}
  \begin{subfigure}{0.49\textwidth}
    \includegraphics[width=9cm]{%
      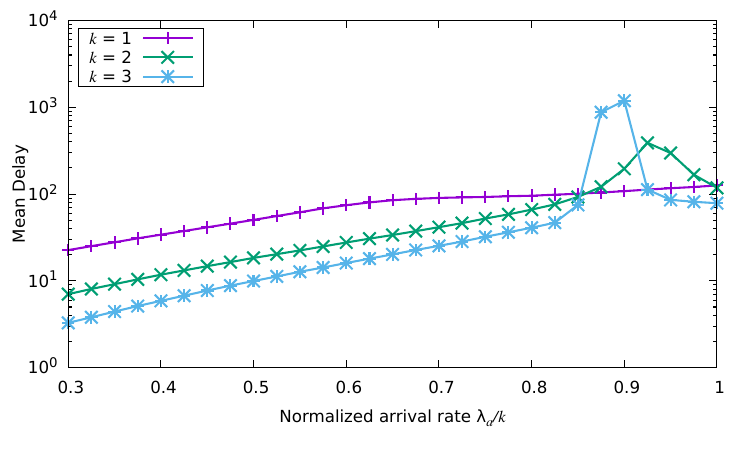}
    \caption{Average delay}
  \end{subfigure}
  \caption{\label{fig:IRSA_n_200} Simulation. Performance for $\Lambda_1(x)$ 
    and $\Lambda_2(x)$, frame length $n = 200$ slots.}
\end{figure}

The $\mathsf{PLR}$ and the average delay are simulated and depicted in
Fig.~\ref{fig:IRSA_n_200}. As the figure shows, the
$\mathsf{PLR}$ is decreasing in $k$ for the same normalized load
$\mathsf{L}$, and the average decoding delay is consistently lower as
$k$ increases when $\mathsf{L}$ is below the asymptotic decoding
threshold, as expected (the decoding delay is measured in number of
slots). There is a phase transition clearly seen around that
threshold, corresponding to the system entering into the waterfall
regime, in which the receiver starts to be unable to decode some
packets, or achieves that only after a long delay. This explains the
increase in decoding delay, which is particularly sharp for $k =
2$. Once $\mathsf{L}$ is beyond the threshold, $\mathsf{PLR}$
increases quickly, so the receiver discards a fraction of the packets
and those which can actually be correctly decoded have a decoding
delay almost independent of $\mathsf{L}$. These results are also
observed for the case without irregular repetition
(Fig.~\ref{fig:IRSA_n_200}, (c)-(d)), with the difference that the end of
the waterfall region occurs at $\mathsf{L} \approx 0.9$ instead of
$\mathsf{L} \approx 0.8$ for the distribution $\Lambda_1(x)$.  We also see
that, below the threshold, the average decoding delay appears to be
linear in $\mathsf{L}$, and that the limiting average decoding delay
when $\mathsf{L} \to 1$ is virtually independent of $k$ and
approximately equal to $n / 2$.

\subsection{PLR and Average Delay with Maximum Delay Constraint}

The last Section showed simulation results for a receiver with
unbounded delay, i.e., all the received replicas are kept into memory
awaiting for a decoding opportunity. This was done only for the
purpose of obtaining the lowest possible $\mathsf{PLR}$, by not
missing any packet decoding event except those due to stopping sets in
the underlying decoding graph. For real systems, $k$-MUD will be
limited either by a maximum delay constraint or by the maximum amount
of memory at the receiver. Thus, we present here the performance under
the first assumption, a maximum delay constraint
$\delta_\text{max}$. The packet loss ratio in this case
$\mathsf{PLR}(\delta_\text{max})$ is defined as the probability that a
packet of an arbitrary users is not successfully resolved within
$\delta_\text{max}$ slots after the user transmits its first replica
packet. Note that the delay constraint does not imply a bound on the
memory used at the receiver, especially for large loads. We will
analyze the setup with finite memory in a subsequent Section.

\begin{figure}[t]
  \centering
  \begin{subfigure}{0.49\textwidth}
    \includegraphics[width=9cm]{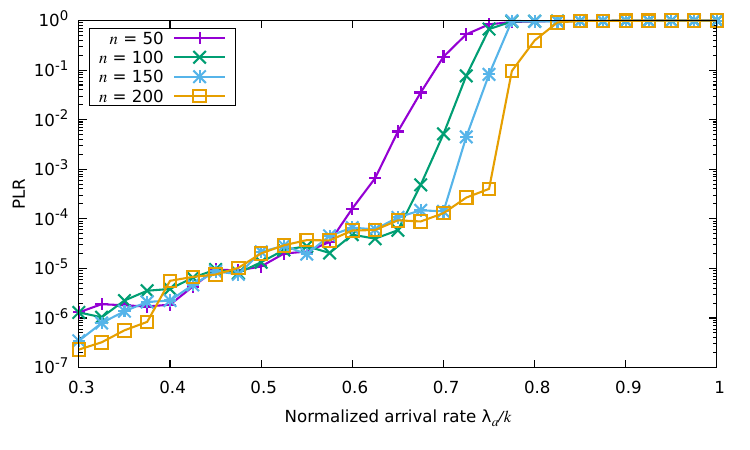}
    \caption{$\mathsf{PLR}(200)$}
  \end{subfigure}
  \begin{subfigure}{0.49\textwidth}
    \includegraphics[width=9cm]{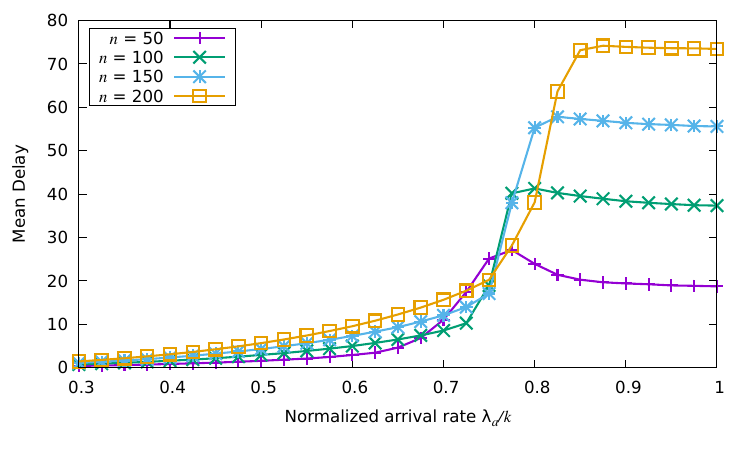}
    \caption{Average delay, $\delta_\text{max} = 200$ slots.}
  \end{subfigure}
  \caption{\label{fig:IRSA_l_3_8:bounded-delay-3}  Performance for
    $3$-MUD IRSA with bounded delay, $\delta_\text{max} = 200$
    slots. Degree distribution $\Lambda_1(x)$.}
\end{figure}
\begin{figure}[tp]
  \centering
  \begin{subfigure}{0.49\textwidth}
    \includegraphics[width=9cm]{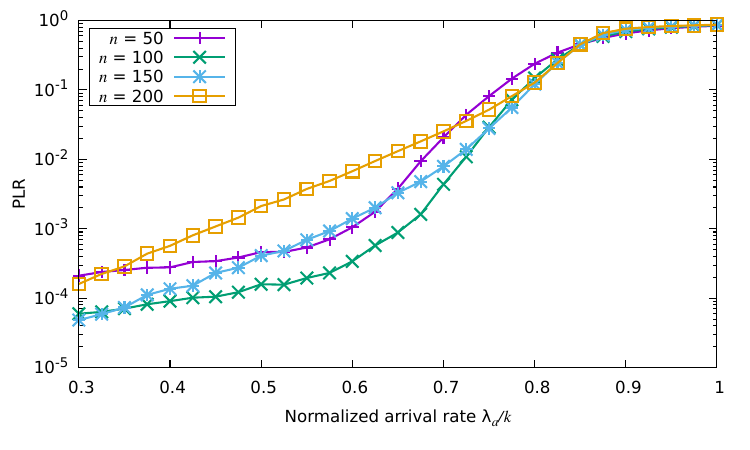}
    \caption{$\mathsf{PLR}(200)$}
  \end{subfigure}
  \begin{subfigure}{0.49\textwidth}
    \includegraphics[width=9cm]{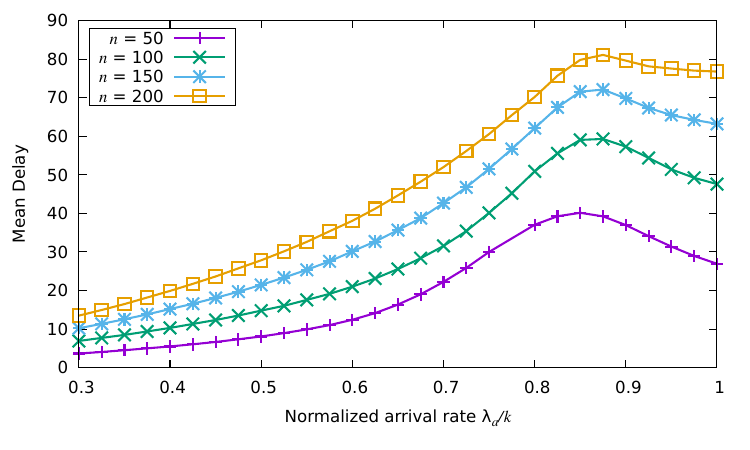}
    \caption{Average delay, $\delta_\text{max} = 200$ slots.}
  \end{subfigure}
  \begin{subfigure}{0.49\textwidth}
    \includegraphics[width=9cm]{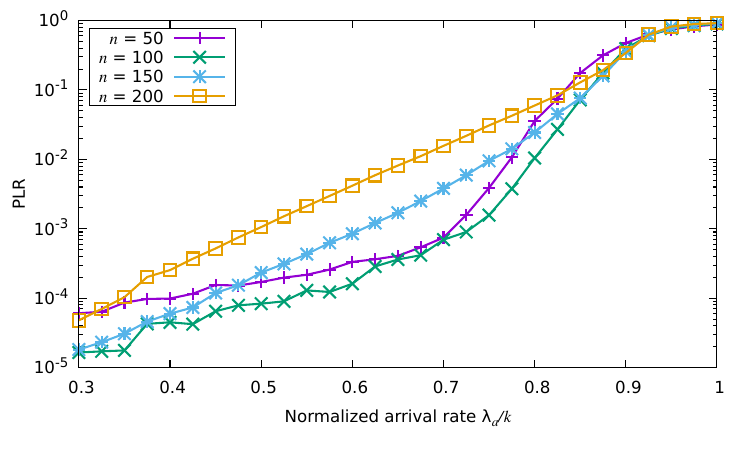}
    \caption{$\mathsf{PLR}(200)$}
  \end{subfigure}
  \begin{subfigure}{0.49\textwidth}
    \includegraphics[width=9cm]{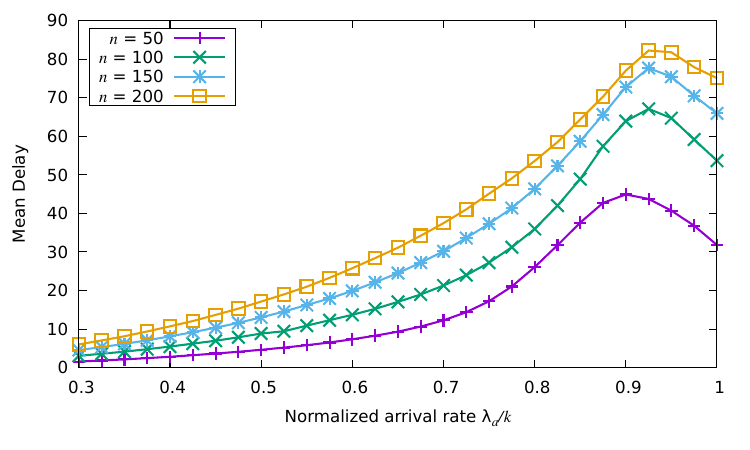}
    \caption{Average delay, $\delta_\text{max} = 200$ slots.}
  \end{subfigure}
  \begin{subfigure}{0.49\textwidth}
    \includegraphics[width=9cm]{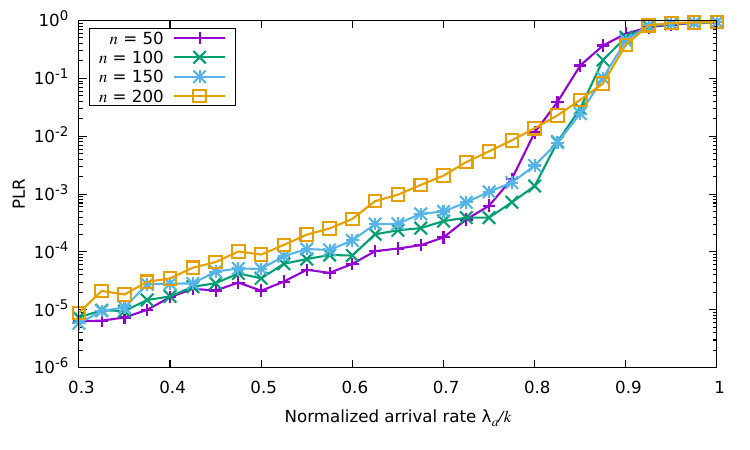}
    \caption{$\mathsf{PLR}(200)$}
  \end{subfigure}
  \begin{subfigure}{0.49\textwidth}
    \includegraphics[width=9cm]{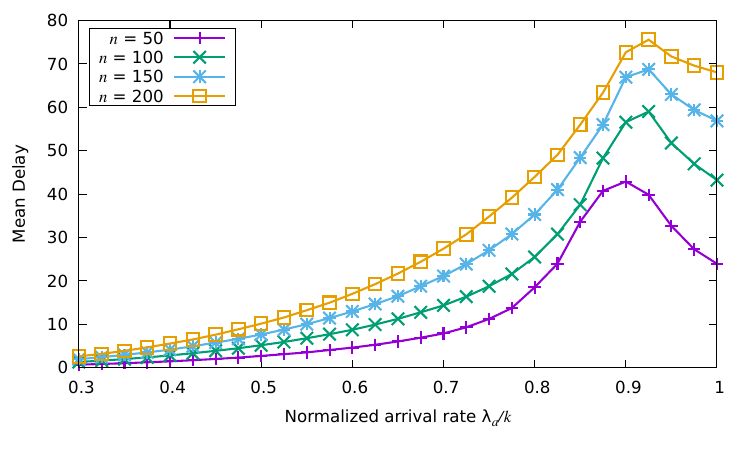}
    \caption{Average delay, $\delta_\text{max} = 200$ slots.}
  \end{subfigure}
  \caption{\label{fig:IRSA_bounded-delay} Performance for IRSA
    with bounded delay, $\delta_\text{max} = 200$ slots. Degree
    distributions $\Lambda_1(x)$ in (a)-(b); $\Lambda_2(x)$ in (c)-(d); $\Lambda_3(x)$ in (e)-(f).}
\end{figure}

The results are shown in
Fig.~\ref{fig:IRSA_l_3_8:bounded-delay-3}. We clearly see that
increasing the frame length under the constraint of maximum delay
shifts the $\mathsf{PLR}$ threshold to the right, as shown in panel
(a), at the cost of a greater average delay (panel (b)). Again, around
the threshold point the delay increases quickly up to its saturation
point, which is in all the experiments under half the frame
length. One of the advantages of multipacket detection is that, for
low or moderate $\mathsf{L}$, the average delay and the $\mathsf{PLR}$
are significantly lower than in the classical collision model
(Fig.~\ref{fig:IRSA_bounded-delay}). These observations also
hold for the case of other distributions, as displayed in
Fig.~\ref{fig:IRSA_bounded-delay}(c)-(d), and (e)-(f).

\begin{figure}[tp]
  \centering
  \begin{subfigure}{0.49\textwidth}
    \includegraphics[width=9cm]{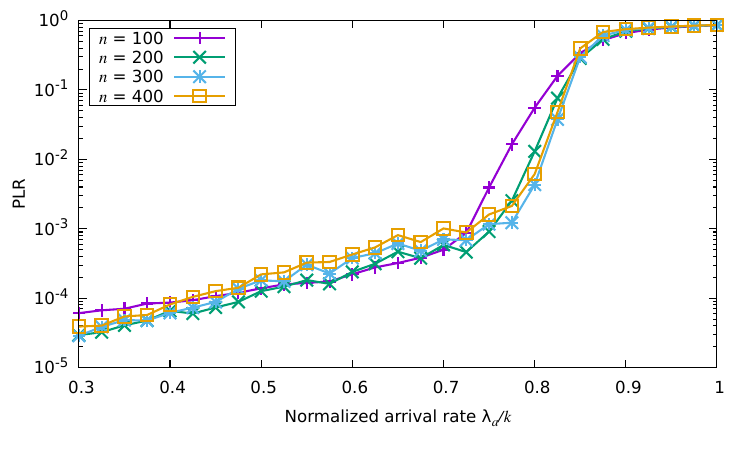}
    \caption{$\mathsf{PLR}$}
  \end{subfigure}
  \begin{subfigure}{0.49\textwidth}
    \includegraphics[width=9cm]{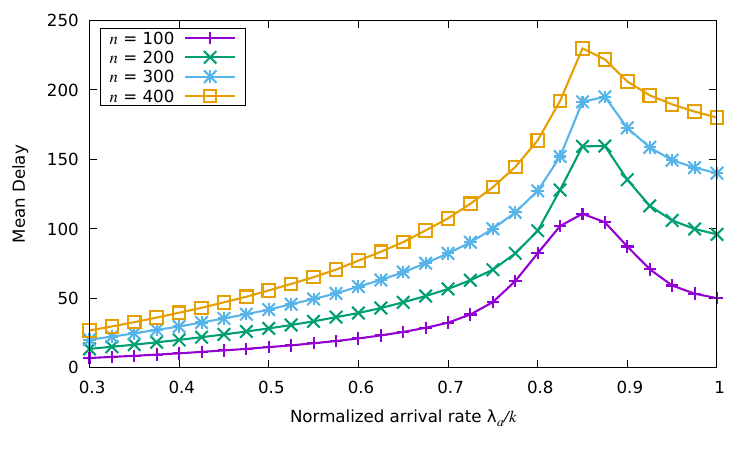}
    \caption{Average delay, $n_\text{max} = 400$ slots.}
  \end{subfigure}
  \begin{subfigure}{0.49\textwidth}
    \includegraphics[width=9cm]{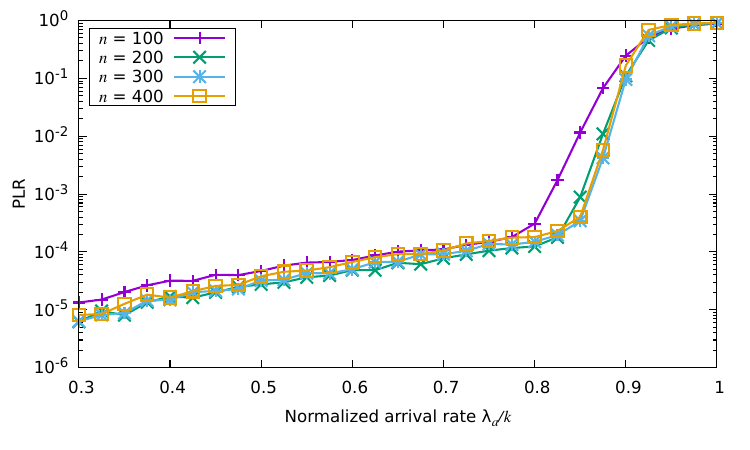}
    \caption{$\mathsf{PLR}$}
  \end{subfigure}
  \begin{subfigure}{0.49\textwidth}
  \includegraphics[width=9cm]{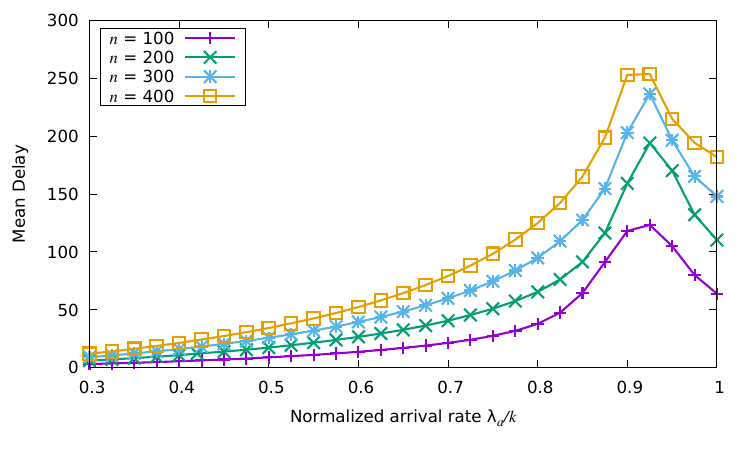}
    \caption{Average delay, $n_\text{max} = 400$ slots.}
  \end{subfigure}
  \begin{subfigure}{0.49\textwidth}
    \includegraphics[width=9cm]{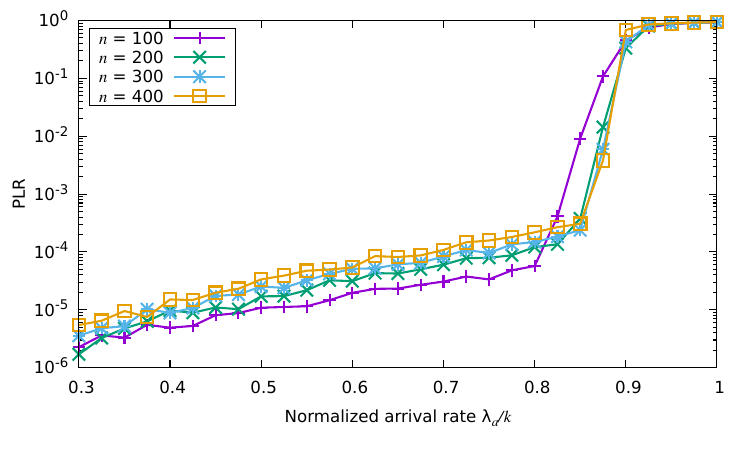}
    \caption{$\mathsf{PLR}$}
  \end{subfigure}
  \begin{subfigure}{0.49\textwidth}
  \includegraphics[width=9cm]{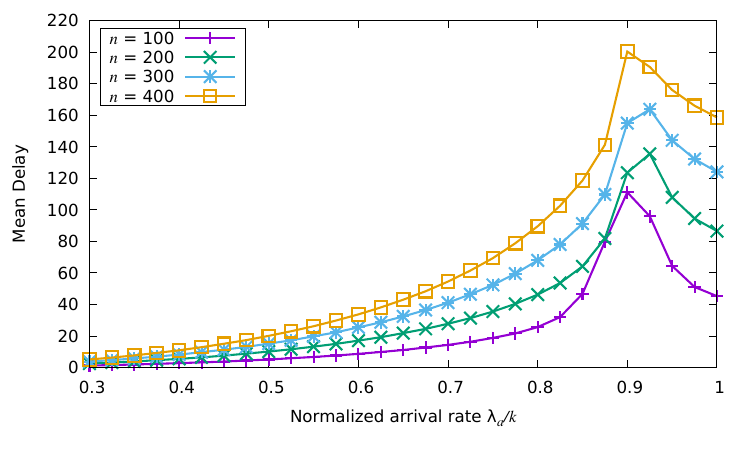}
    \caption{Average delay, $n_\text{max} = 400$ slots.}
  \end{subfigure}
  \caption{\label{fig:IRSA_l_3_8:bounded-memory-1}  Performance for
    IRSA with bounded memory, $n_\text{max} = 400$ slots. Degree
    distributions $\Lambda_1(x)$ in (a)-(b); $\Lambda_2(x)$ in (c)-(d); $\Lambda_3(x)$ 
    in (e)-(f).}
\end{figure}

\subsection{PLR Loss Rate and Mean Delay with Maximum Memory
  Constraint}

Instead of enforcing a maximum delay, a more natural constraint to
reduce the complexity of the receiver is to limit the amount of memory
used to hold the received packets while awaiting for a decoding
opportunity. A small memory size is desirable, but it might lead to
miss the decoding of some packets when the traffic load is high. To
explore this trade-off, we have simulated the system using a fixed
amount of memory $n_\text{max} = 400$ packets (we consider each
replica a different packet for counting the memory
use). Figure~\ref{fig:IRSA_l_3_8:bounded-memory-1} collects the results for
comparison of pure IRSA ($k = 1$, top), $2$-MUD (middle), and $3$-MUD, $2$-MUD (bottom) 
in this respect.

\section{Conclusions}
\label{sec:conclusions}

Introducing the capability of multipacket reception into uncoordinated
wireless communications improves the system utilization and its energy
efficiency, simultaneously, contributing to a scalable solution able
to support tens to hundreds of devices, as in IoT and mMTC
applications. We derived in this paper the dynamic density evolution
equations for performance analysis or IRSA with $k$-MUD at the SIC
receiver(s). Unlike previous works in the literature, 
where the analysis focused on an asymptotically large number of slots, 
our results hold for the finite length regime.
The expressions are tractable due to the memoryless
property of the Poisson distribution, and involve only modified terms
to account for the effective decoding probability at each time slot at
the receiver. This density evolution approach can be used to calculate
the associated potential function and the decoding threshold for IRSA,
and provides a benchmark for numerical approximations to the error
floor probability like those developed
in~\cite{Liva2011,Clazzer2022}. Our simulation results quantify the
effects of MUD, finite memory and/or bounded maximum delay at the
decoder, and show that MUD does not require, in general, long frame
lengths nor excessive memory to perform well up to high normalized
traffic loads. These observations emphasize the importance of
introducing SIC in IRSA receivers, even to a low degree. The present
work can be extended in several directions. One obvious direction is
to consider more realistic model for SIC and analyze the robustness of
the schemes against channel impairments, power control, or CSIT
estimation. Another possibility is to use the framework to optimize
the degree distribution of IRSA in terms of signal-to-noise ratio,
offered load, and rate.

\section{Acknowledgments}
This work was supported by the Spanish Government under grants ``Enhancing Communication Protocols with Machine Learning while Protecting Sensitive Data (COMPROMISE)" PID2020-113795RB-C33, and ICARUS ``Red satelital centrada en información para comunicaciones vehiculares'' PID2020-113240RB-I00, funded by MCIN/AEI/10.13039/501100011033.

\appendix

\section{Proof of Theorem~\ref{thm:de}}
\label{app:proof-a}


The proof is divided in two cases. The system with fixed transmission
in the first slot of the user's virtual frame is more involved and is
addressed first. The proof technique is the same for uniform
distribution of the packet replicas in IRSA. The key property is that
the messages that arrive at a user or receiver node are a thinned
Poisson process that depends on the previous density values in a
sliding window of length $n$ and on the degree distributions.

\subsection{Density evolution with transmission in the first slot}

Suppose the first replica in sent always in the first slot after its
arrival to the system. Then, a variable node at epoch $i$ is always
connected to the check node at epoch $i$, while the remaining edges
are connected to other future positions randomly chosen. Thus, there
are two different edge types and the density evolution involves the
quantities $p_{(i,i)}$, $p_{(i,j)}$, $q_{(i,i)}$ and $q_{(i,j)}$ for
the probabilities of the messages exchanged between user nodes and
receiver nodes (see again Figure~\ref{fig:IRSA-example1}). The
subscripts make clear the orientation and type of the edges.

Let us derive first $p_{(i,i)}$ and $p_{(i,j)}$, the probabilities of
decoding failures in edges $(i, i)$ and $(i, j)$, seen from a user
node. An outgoing message from user node $i$ fails if all of its
incoming messages also fail, i.e., if there is no slot such that the
replica can be successfully decoded. Suppose the user node has degree
$\mathsf{r}$ and is at position $i$, so it has an edge to the receiver
node at position $i$ and other $\mathsf{r} - 1$ edges at positions
beyond $i$. Since the edges that connect the user node to the receiver
nodes are drawn uniformly from the interval $[i + 1, i + n - 1]$ the
average incoming failure probability along each edge is the same, and
is given by
\begin{equation}
  \label{eq:average-qii}
  \overline{q}_{(i, i)} := \frac{1}{n - 1} \sum_{j = i + 1}^{i + n - 1} q_{(j,i)}.
\end{equation}
Thus, the probability that the message sent along $(i, i)$ is an
erasure is $\overline{q}_{(i, i)}^{\mathsf{r} - 1}$, and averaging over
the degree distribution $\Lambda^{(i)}(x)$, we get
\begin{equation}
  \label{eq:average-pii}
  p_{(i, i)} = \sum_{\mathsf{r}} \Lambda^{(i)}_{\mathsf{r}} \overline{q}_{(i,
    i)}^{\mathsf{r} - 1}  := \Lambda^{(i)}\bigl( \overline{q}_{(i,i)}
  \bigr). 
\end{equation}
In an analogous way, $p_{(i, j)}$ can be obtained as
\begin{equation}
  \label{eq:average-pij}
  p_{(i, j)} = q_{(i, i)} \sum_{\mathsf{r}} \lambda_{\mathsf{r}}^{(i)}
  \overline{q}_{(i, i)}^{\mathsf{r} - 2} = q_{(i, i)} \lambda^{(i)}(\overline{q}_{(i, i)})
\end{equation}
for $j \in [i + 1, i + n - 1]$.

The derivation of the probabilities $q_{(i, i)}$ and $q_{(i, j)}$ can
be done as follows. With $k$-MUD, an edge outgoing from receiver node
$i$ is a decoding failure if at least $k$ of its incoming edges
(distinct from the outgoing edges) fail. This is because the receiver
can decode perfectly any slot with $k$ or less packets, so only slots
with more than $k$ packets are unsuccessful. Conversely, a message
from a receiver node is not a failure if less than $k$ of its other
$r_1 + r_2 - 1$ incoming edges are erasures. Consider a receiver node
of degree $\mathsf{r}$ at position $i$, say $w$. Node $w$ has $r_1$
edges connected to type $i$ user nodes, and $r_2$ edges connected
to nodes in the time window
\begin{equation}
  \mathcal{J}_i =\begin{cases}
    \emptyset, & \quad \text{for $i = 1$} \\
    [1, i - 1], & \quad 2 \leq i < n \\
    [i - n + 1, i - 1], & \quad i \geq n.
  \end{cases}
\end{equation}

Then the (conditional) probability that an outgoing message along an
edge $e$ connecting $w$ to a variable node of type $i$ is a failure,
$p_e$, is the probability that $k_1$ of the remaining $r_1 - 1$ edges
connecting $w$ to type $i$ variable nodes, and $k_2$ out of the $r_2$
edges connecting $w$ to some variable node in $\mathcal{J}_i$ are all
failures, for any pair $(k_1, k_2)$ such that $k_1 + k_2 \geq
k$. Hence

\begin{equation}
  \label{eq:erasure-prob}
  p_e = \sum_{k_1 + k_2 \geq k} \binom{r_1 - 1}{k_1} p_{(i, i)}^{k_1}
  (1 - p_{(i, i)})^{r_1 - 1 - k_1} \binom{r_2}{k_2} \overline{p}_i^{k_2} (1 -
  \overline{p}_i)^{r_2 - k_2} 
\end{equation}
where it must be understood that $\binom{a}{b} = 0$ if $a < b$. The
particular case $k = 1$, i.e., without MUD, gives
$ 1 - p_e = (1 - p_{(i, i)})^{r_1 - 1} (1 - \overline{p}_i)^{r_2}$.
In the above formulas
\begin{equation}
  \label{eq:average-pi}
  \overline{p}_i = \begin{cases}
    0, & \quad\text{for $i = 1$} \\
    \sum_{k \in \mathcal{J}_i} p_{(k, i)} / (i - 1), & \quad\text{for $1 < i
      <n$} \\
    \sum_{k \in \mathcal{J}_i} p_{(k, i)} / (n - 1), & \quad\text{for $i \geq n$}
  \end{cases}
\end{equation}
is the average failure probability of incoming messages from user
nodes in the time interval $\mathcal{J}_i$.  Now, we
average~\eqref{eq:erasure-prob} over the edge-perspective receiver
node distribution and the node-perspective receiver node distribution,
jointly. We have

\begin{align}
  q_{(i, i)} &= \Esp_{r_1 \sim \gamma^{(i)}} \Esp_{r_2 \sim \Gamma^{(i)}} \left[
      \sum_{k_1 + k_2 \geq k} \binom{r_1 - 1}{k_1} p_{(i, i)}^{k_1} (1
      - p_{(i, i)})^{r_1 - 1 - k_1} \binom{r_2}{k_2} \overline{p}_i^{k_2}
      (1 - \overline{p}_i)^{r_2 - k_2} \right]
    \label{eq:expectation-qii-a} \\
    &= \sum_{r_1, r_2}
      \Prob(\gamma^{(i)} = r_1) \Prob(\Gamma^{(i)} = r_2) \cdot \notag \\
             &
     \sum_{k_1 + k_2 \geq k} \binom{r_1 -
        1}{k_1} p_{(i, i)}^{k_1} (1 - p_{(i, i)})^{r_1 - 1 - k_1} \\
      & \binom{r_2}{k_2} \overline{p}_i^{k_2} (1 - \overline{p}_i)^{r_2 -
        k_2}. \label{eq:expectation-qii-b}
\end{align}

Interchanging the outer and inner summations this can be rearranged as

\begin{multline}
  \label{eq:expectation-qii}
  q_{(i, i)} = \sum_{k_1 + k_2 \geq k} \left( \sum_{r_1 = 1}^\infty
  \Prob(\gamma^{(i)} = r_1) \binom{r_1 - 1}{k_1} p_{(i, i)}^{k_1} (1 - p_{(i,
    i})^{r_1 - 1 - k_1} \right) \cdot \\
\left(
  \sum_{r_2 = 0}^\infty \Prob(\Gamma^{(i)} = r_2) \binom{r_2}{k_2}
  \overline{p}_i^{k_2} (1 - \overline{p}_i)^{r_2 - k_2} \right). 
\end{multline}

Now, for the first sum

\begin{align}
  &\sum_{r_1 = 1}^\infty \Prob(\gamma^{(i)} = r_1) \binom{r_1 - 1}{k_1} p_{(i,i)}^{k_1} (1 - p_{(i, i)})^{r_1 - 1 - k_1} \notag \\
  &=\sum_{r_1 =  0}^\infty \exp(-\lambda_a) \frac{\lambda_a^{r_1}}{r_1!}
  \binom{r_1}{k_1} p_{(i,i)}^{k_1} (1 - p_{(i, i)})^{r_1 - k_1}
  \label{eq:partial-a} \\ 
  &= \frac{(\lambda_a p_{(i, i)})^{k_1}}{k_1!} \exp(-\lambda_a) \sum_{r_1 =
    0}^\infty \frac{(\lambda_a (1 - p_{(i, i)}))^{r_1}}{r_1!} \notag \\
  &=\exp(-\lambda_a p_{(i, i)}) \frac{(\lambda_a p_{(i,
      i)})^{k_1}}{k_1!} \label{eq:partial-b}
\end{align}

where equality~\eqref{eq:partial-a} follows from the fact that
$\gamma_{r_1}^{(i)} = \Gamma_{r_1 - 1}^{(i)}$ and the arrivals are
Poissonian, and~\eqref{eq:partial-b} follows because
$\sum_{n = 0}^\infty x^n / n! = \exp(x)$.  For the second term,
using Theorem~\ref{th:degree-dist} and the fact that $\Gamma^{(i)}$
follows a Poisson distribution
\begin{multline}
  \label{eq:partial2}
  \sum_{r_2 = 0}^\infty \Prob(\Gamma^{(i)} = r_2) \binom{r_2}{k_2}
  \overline{p}_i^{k_2} (1 - \overline{p}_i)^{r_2 - k_2} \\
  = \sum_{r_2 = 0}^\infty
  \binom{r_2}{k_2} \overline{p}_i^{k_2} (1 - \overline{p}_i)^{r_2 - k_2}
  \exp(-z \delta_i) \frac{(z \delta_i)^{r_2}}{r_2!} \\
  = \frac{(z \overline{p}_i \delta_i)^{k_2}}{k_2!} \exp(-z \overline{p}_i \delta_i)
\end{multline}
where $z = (\Lambda^\prime(1) - 1) / (n - 1)$.
Using~\eqref{eq:partial-b} and~\eqref{eq:partial2}
in~\eqref{eq:expectation-qii} we have
\begin{equation}
  q_{(i, i)} =  \sum_{k_1 + k_2 \geq k} \exp(-\lambda_a p_{(i, i)} - z
  \tilde{p}_i \delta_i) \frac{(\lambda_a p_{(i, i)})^{k_1}}{k_1!}
  \frac{(z \overline{p}_i \delta_i)^{k_2}}{k_2!}  = \Prob(V_1  + V_2 \geq k)
\end{equation}
where $V_1 \sim \mathsf{Poisson}(\lambda_a p_{(i,i)})$,
$V_2 \sim \mathsf{Poisson}(z \overline{p}_i \delta_i)$ and $V_1$, $V_2$ are
independent random variables. Since the sum of independent Poisson
variables is Poisson, we finally conclude
\begin{equation}
  \label{eq:closed-form-qii}
  q_{(i, i)} = 1 - \sum_{m = 0}^{k - 1} \exp(-\nu_i) \frac{\nu_i^m}{m!}
\end{equation}
with $\nu_i \triangleq \lambda_a p_{(i, i)} + z \overline{p}_i \delta_i
= \lambda_a p_{(i, i)} + \bigl( \Lambda^\prime(1) - 1) / (n - 1)
\overline{p}_i \delta_i$. \\

In a similar way, the probability $q_{(i, j)}$ can be derived after
averaging

\begin{equation}
  \label{eq:expectation-qij}
  \sum_{k_1 + k_2 \geq k} \binom{r_1}{k_1} p_{(i, i)}^{k_1}
  (1 - p_{(i, i)})^{r_1 - k_1} \binom{r_2 - 1}{k_2}  \overline{p}_i^{k_2} (1 -
  \overline{p}_i)^{r_2 - 1 - k_2}
\end{equation}

over the node-perspective and the edge-perspective receiver node degree
distributions, i.e.
\begin{align}
  q_{(i, j)} &= \Esp_{r_2 \sim \gamma^{(i)}} \Esp_{r_1 \sim \Gamma^{(i)}} \left[
        \sum_{k_1 + k_2 \geq k} \binom{r_1}{k_1} p_{(i, i)}^{k_1} (1 -
        p_{(i, i)})^{r_1 - k_1} \binom{r_2 - 1}{k_2} \overline{p}_i^{k_2} (1 -
        \tilde{p}_i)^{r_2 - 1 - k_2} \right]
      \label{eq:expectation-qij-a} \\
      &= \sum_{r_1 = 0}^\infty \sum_{r_2 = 1}^\infty
      \Prob(\gamma^{(i)} = r_2) \Prob(\Gamma^{(i)} = r_1) \cdot \notag \\
      &\left( \sum_{k_1 + k_2 \geq k}
        \binom{r_1}{k_1} p_{(i, i)}^{k_1} (1 - p_{(i, i)})^{r_1 - k_1} \binom{r_2 - 1}{k_2} \tilde{p}_i^{k_2} (1 - \tilde{p}_i)^{r_2 - 1 -
          k_2} \right). \label{eq:expectation-qij-b}
\end{align}
Noting that~\eqref{eq:erasure-prob} and~\eqref{eq:expectation-qij}, and
also~\eqref{eq:expectation-qii-a} and~\eqref{eq:expectation-qij-a} are
the same if we swap indices $1$ and $2$, we obtain
\begin{equation}
  \label{eq:density5}
  q_{(i, j)} = q_{(i, i)} = 1 - \sum_{m = 0}^{k - 1} \exp(-\nu_i) \frac{\nu_i^m}{m!}.
\end{equation} 
Equality follows from the fact that
$\Gamma^{(i)}(x) = \gamma^{(i)}(x)$ which is in turn a consequence of
the underlying Poisson distributions. This proves part (i) in
Theorem~\ref{thm:de}. Notice that~\eqref{eq:density5} is independent
of $j$.

\subsection{Density evolution with uniformly distributed replicas}

When the slots for transmitting the replicas are chosen uniformly
within the local frame there is no need to consider different types of
edges. Therefore, the probability $p_{i}$ that a failure message is
passed from a variable node at position $i$ is
\begin{equation}
  \label{eq:de-fa-csa-u1}
  p_{i} = \sum_{\mathsf{r}} \lambda_{\mathsf{r}}
  \overline{q}_i^{\mathsf{r} - 1} = \lambda(\overline{q}_i) 
\end{equation} 
where
\begin{equation}
  \label{eq:de-fa-csa-u2}
  \overline{q}_i = \frac{1}{n} \sum_{j = i}^{i + n - 1} q_j
\end{equation}
is the average probability of the incoming edges to the variable node at
position $i$. Similarly, it is possible to repeat the analysis in the previous Section for deriving that the probability that a check node at position $i$ sends a failure message is
\begin{equation}
  \label{eq:de-fa-csa-u3}
  q_i = \sum_r \gamma_{i,r} \sum_{j \geq k} \binom{r - 1}{j} \overline{p}_i^j (1 -
  \overline{p}_i)^{r - 1 - j} 
  = 1 - \sum_{j = 0}^{k - 1} \exp(-\eta_i) \frac{\eta_i^j}{j!}
\end{equation}
where $\eta_i \triangleq \mu_i \overline{p}_i \Lambda^\prime(1)/ n$ and
\begin{equation}
  \label{eq:de-fa-csa-u4}
  \overline{p}_i = \begin{cases}
    \sum_{j = 1}^i \frac{p_j}{i}, \quad & 1 \leq i < n \\
    \sum_{j = i - n + 1}^i \frac{p_j}{n}, \quad & i \geq n.
  \end{cases}
\end{equation}
This completes the proof of the second statement in
Theorem~\ref{thm:de}. The threshold $\lambda_a^\star$ is
found by searching for the largest value of $\lambda_a$ for which
$\overline{p}_i$ converges to $0$ for all positions.

\section{Proof of Lemma~\ref{lemma:maximum}}
\label{app:proof-b}

Define the auxiliary functions $g_k(x) = 1 - \exp(-\zeta x) \sum_{j =
  0}^{k - 1} (\zeta x)^j / j!$ and $h_k(x) = \sum_{j = 0}^{k - 1}
(\zeta x)^j / j! (\zeta x + k - j)$. Then, it is a straightforward yet
long calculation to check that
\begin{equation}
  \label{eq:u-prime}
  U^\prime_k(x) = \exp(-\zeta x) \left( h_k(x) -
    \frac{h^\prime_k(x)}{\zeta} - \frac{\Lambda^\prime\bigl( g_k(x)
      \bigr)}{\Lambda^\prime(1)}  \zeta \frac{(\zeta x)^{k - 1}}{(k - 1)!} \right).
\end{equation}
Using the identity

\begin{equation}
  h_k^\prime(x) = \zeta \sum_{j = 0}^{k - 1} \frac{(\zeta x)^{j -
      1}}{j!} + h_{k-1}(x),
\end{equation}

plugging this into~\eqref{eq:u-prime}, and simplifying terms we get
\begin{equation}
  U^\prime_k(x) = \exp(-\zeta x) \zeta \frac{(\zeta x)^{k - 1}}{(k - 1)!}
  \left( x - \lambda \bigl( g_k(x) \bigr) \right),
\end{equation}
so the maximum of $U(x)$ is attained
when the term between parentheses is zero. This proves the lemma.

\bibliographystyle{elsarticle-num}
\bibliography{MPR-IRSA}

\end{document}